\documentclass[journal]{IEEEtran}

\usepackage[T1]{fontenc}% optional T1 font encoding
\usepackage{amsmath}
\usepackage{amsfonts}
\usepackage{color}
\usepackage{soul}
\usepackage{graphicx}
\usepackage[caption=false,font=footnotesize]{subfig}
\usepackage{tikz}
\usetikzlibrary{positioning}
\usetikzlibrary{automata,arrows,positioning,calc}
\usepackage{cite}
\usepackage{comment}

\usepackage{afterpage,array,rotating}
\usepackage{comment}

\usepackage{algorithm}
\usepackage{algorithmic}
\usepackage{array}
\usepackage{multirow}
\usepackage{tikzpagenodes}
\usepackage{lipsum}

\normalsize
\begin{document}

\title{IRS-aided Near-field Communication: \\Prospects and Challenges with Codebook Approach}

\author{
	\IEEEauthorblockN{
		Ryuhei Hibi$^{\dag}$$^1$,~\IEEEmembership{Graduate Student Member,~IEEE,}
        Hiroaki Hashida$^{\dag}$$^{\ddag}$$^2$,~\IEEEmembership{Member,~IEEE,}\\
		Yuichi Kawamoto$^{\dag}$$^3$,~\IEEEmembership{Member,~IEEE,}
		Nei Kato$^{\dag}$$^4$,~\IEEEmembership{Fellow,~IEEE,}\\
	}
	\IEEEauthorblockA{
		$^{\dag}$Graduate School of Information Sciences, Tohoku University, Sendai, Japan\\
        $^{\ddag}$Frontier Research Institute for Interdisciplinary Sciences, Tohoku University, Sendai, Japan\\
		E-mails: ryuhei.hibi.p1@dc.tohoku.ac.jp$^1$, \\
                \{hiroaki.hashida.d6$^2$, yuichi.kawamoto.d4$^3$, nei.kato.d3$^4$\}@tohoku.ac.jp
	}
}

% The paper headers
\markboth{IEEE Wireless Communications}%
{Submitted paper}

% make the title area
\maketitle

\begin{comment}
\begin{tikzpicture}[remember picture,overlay]
    \node[scale=0.6,align=center,text=black] at ([yshift=-3em]current page text area.south) {
    Copyright (c) 2022 IEEE. Personal use of this material is permitted. However, permission to use this material for any other purposes must be obtained from the IEEE by sending a request to pubs-permissions@ieee.org.};
\end{tikzpicture}%
\end{comment}

\begin{abstract}
Intelligent reflecting surfaces (IRSs) are gaining attention as a low-cost solution to the coverage reduction in high-frequency bands used in next-generation communications. IRSs achieve low costs by controlling only the reflection of radio waves. However, to improve further the propagation environment, larger IRS sizes are required owing to their inability to amplify and retransmit signals. As the IRS size increases, the near-field region expands, requiring beamfocusing instead of beamforming, which is extensively used in existing research. This results in considerable overhead for IRS control decisions. To address this, constructing a codebook that achieves high communication quality with fewer IRS control patterns is effective. This article presents experimental results demonstrating the effectiveness of beamfocusing, construction policy for nonuniform three-dimensional codebooks, and simulation evaluation results of communication performance when operating IRSs with various codebooks. We believe these insights will foster further value for IRSs in next-generation communications.
\end{abstract}
%\end{comment}

% Note that keywords are not normally used for peer-reviewed papers.

\IEEEpeerreviewmaketitle

\section{Introduction}
\label{section:Introduction}
\IEEEPARstart{A}{s} the information society continues to evolve, the demand for wireless communication has steadily increased~\cite{Demand_Chief}, which has led to restrictions on the frequency bands available for communication. Consequently, efforts are underway to use higher frequency bands, such as millimeter waves, as new frequency resources for communication. However, higher frequencies have the inherent property of greater distance attenuation and increased directivity, which reduces the coverage area that a single base station can provide. This presents a challenge as the number of base stations required to provide comprehensive coverage becomes enormous, making the widespread adoption of high-frequency bands a cost-prohibitive bottleneck.
One approach used to address this issue is the use of intelligent reflecting surfaces (IRSs) that achieve low costs by controlling only the reflection gain of radio waves. IRSs are devices that integrate on a plane surface metamaterial elements capable of dynamically controlling the reflection phase of incident radio waves~\cite{IRS_Chief}. By coordinating the operation of these elements, the gain can be increased by coherently adding radio waves at arbitrary points or suppressing interference based on antiphase addition. By forming reflection paths that bypass obstructions and compensating for distance attenuation through directional control using IRSs, the coverage area of a base station can be expanded. Moreover, as IRSs only reflect radio waves, they offer significantly higher cost-effectiveness in terms of manufacturing, installation, and management compared with the installation of repeaters or base stations that retransmit received signals, thus addressing the bottleneck in the adoption of high-frequency band communications~\cite{kato_1, kato_2}.

While IRSs achieve low costs by not amplifying or retransmitting radio waves, their sizes heavily affects the extent to which they can expand the communication ranges. This is because the IRSs only control the reflection of radio waves and are constrained by the power received on the surface. In addition, the directional control by IRSs provide higher resolution and sharper, farther-reaching reflected waves as the number of elements increases. Therefore, to enhance the effects of eliminating dead zones through IRSs, scaling up the size of the IRSs is crucial. Given their thin form factor, the IRSs are envisioned to be installed on building walls, making their large-scale deployment evident~\cite{IRS:Large}. However, the scaling up of IRSs increases the plane-wave approximation error; this risks rendering many existing studies ineffective owing to the manner in which they simplify the problem by assuming perfect plane-wave knowledge. The region wherein a) the plane-wave approximation error is large and b) the propagation characteristics resemble spherical waves is referred to as the near-field; accordingly, increasing the size of the IRSs correspondingly expands the near-field~\cite{NF:Introduction}. Therefore, establishing control methods for supporting users distributed in the near-field becomes crucial for large-scale IRSs. However, near-field research has predominantly been limited to the examination of transmitters and receivers, such as array antennas, with few studies considering the characteristics of IRSs that control only phase rotation. Furthermore, these studies often assume complete possession of channel information for control, with minimal exploration of beamtraining-based IRS control that can account for control overhead.

\begin{figure*}[t]
	\centering
	\includegraphics[width=1.0\hsize]{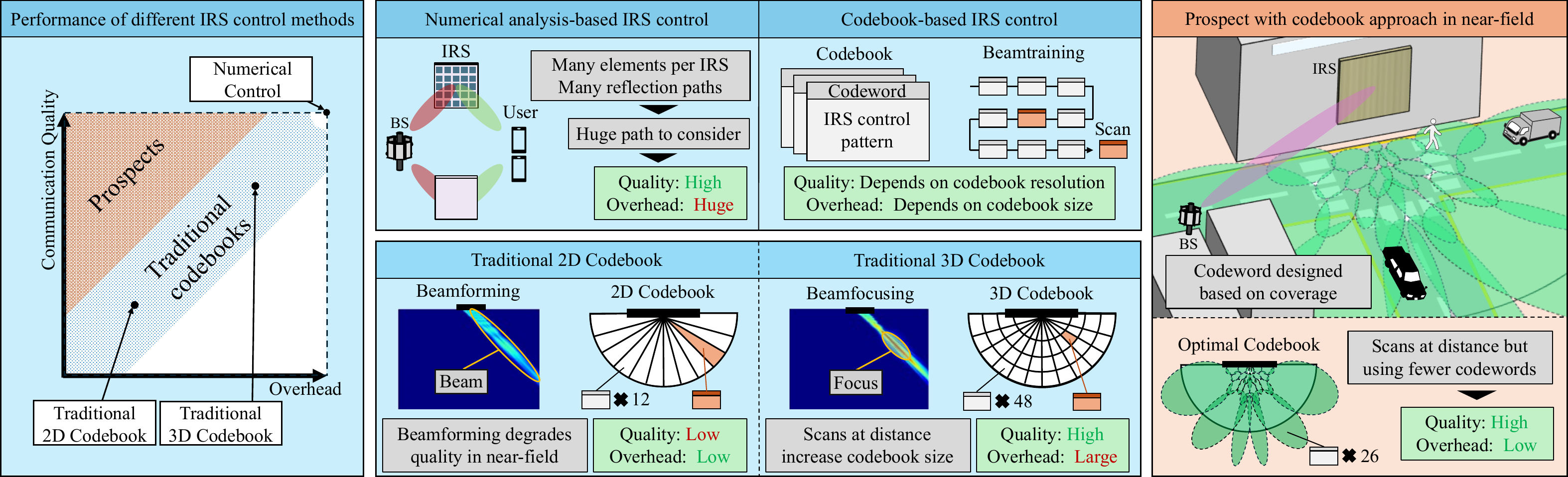}
	\caption{Traditional and prospects of intelligent reflecting surface (IRS) control methods in the near-field.}
	\label{fig:Introduction}
\end{figure*}

As shown in the numerical analysis-based IRS control of Fig.~\ref{fig:Introduction}, in near-field IRS control, the propagation paths that must be considered become enormous, rendering it difficult to measure all channel states and make control decisions based on numerical calculations. Therefore, as shown in the codebook-based IRS control of Fig~\ref{fig:Introduction}, a beamtraining method using a codebook that prepares several IRS controls in advance and scans them for IRS control decisions is considered effective. However, in the near-field, the accuracy of the plane-wave assumption diminishes, resulting in degraded beamforming performance when using a two-dimensional (2D) codebook that considers only the reflection direction as used in the far-field. Conversely, beamfocusing, adopted in a three-dimensional (3D) codebook---effective for the near-field---also requires scanning in the distance direction, which increases the codebook size, and consequently causes an increase in the overhead of beamtraining. Therefore, this paper discusses the main challenges and solutions for associated with the design of a codebook that provides high communication quality in the near-field with a small codebook. The insights revealed through the discussion, as shown in Fig~\ref{fig:Introduction}, are constructed nonuniformly according to the shape of the formed beam. This approach mitigates substantial performance degradation in the near-field while reducing the number of codewords set wastefully despite minimal performance improvement, thereby improving overhead. Furthermore, simulation results used to verify the performance of the proposed codebook are provided. In the following sections, the details of this overview are explained based on the results of practical experiments and simulations. 

\begin{comment}
The structure of this article is as follows: first, we clarify the challenges associated with the expansion of the near-field due to the scaling up of IRSs, which constitutes the focus of this article. We then organize the concept of codebooks used in beamtraining for IRS control decisions and discuss the type of codebooks that should be adopted for controlling large-scale IRSs. We also explain the important characteristics that must be considered and propose a construction method for nonuniform, 3D codebooks effective for controlling large-scale IRSs. Additionally, we conduct simulation evaluations of the performance of various codebooks in near-field communication to demonstrate the effectiveness of the proposed approach. Finally, we summarize the article and present future issues to be addressed. 
\end{comment}

\section{Challenges of IRS-aided Near-field Communications}
\label{section:Charanges}
In beamforming control based on the plane-wave approximation for IRS, the phase rotation necessary for the coherent addition of the radio waves propagating through the surface elements is calculated by assuming that the distances from each element to the source and destination of the signal being reflected depend only on the angles of the transmission and reception points from the center position of the IRS~\cite{Beamforming}. In contrast, beamfocusing control, which is suitable for the near-field, calculates the phase rotation based on the exact distance between the actual position of each surface element and the actual positions of the source and destination of the reflected signal~\cite{Beamfocusing}. An adequately large distance between the IRS and the source/destination allows optimal control through beamforming. However, in the near-field, errors in the plane-wave approximation cause deviations from optimal control. The Fraunhofer distance---often used as the range for the near-field---is defined as $2D^2/\lambda$, where $D$ is the aperture size, and $\lambda$ is the wavelength. Thus, the size of the near-field is larger for larger IRSs, heightening the need for beamfocusing. However, many existing studies still use beamforming for IRS control, meaning that IRS control decisions that are more suitable for near-field communications are required.

In this section, we experimentally demonstrate that beamfocusing is better than beamforming in the near-field before discussing the drawbacks in terms of overhead for beamfocusing.

\subsection{Experimental Study}
\label{subsection:Experiment}
\begin{figure}[t]
	\centering
	\includegraphics[width=1.0\hsize]{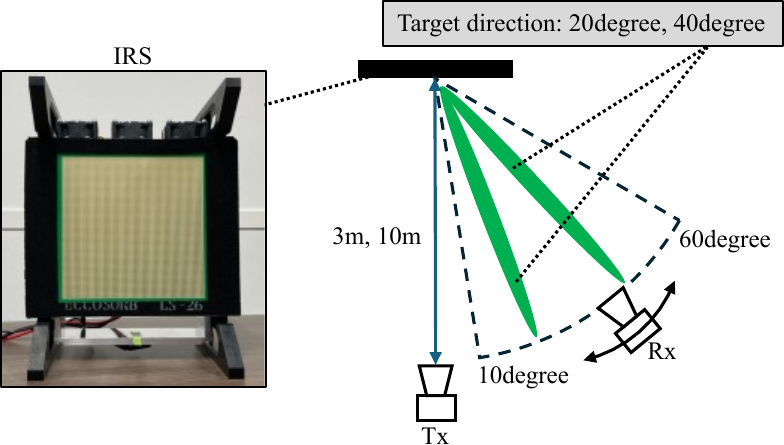}
	\caption{Experimental environment for beamforming and beamfocusing by IRS.}
	\label{fig:ExEnv}
\end{figure}

\begin{figure}[t]
    \centering
    \subfloat[Experimental results when the distance from source/destination to IRS is set to 3 m]{
        \includegraphics[width=0.9\hsize]{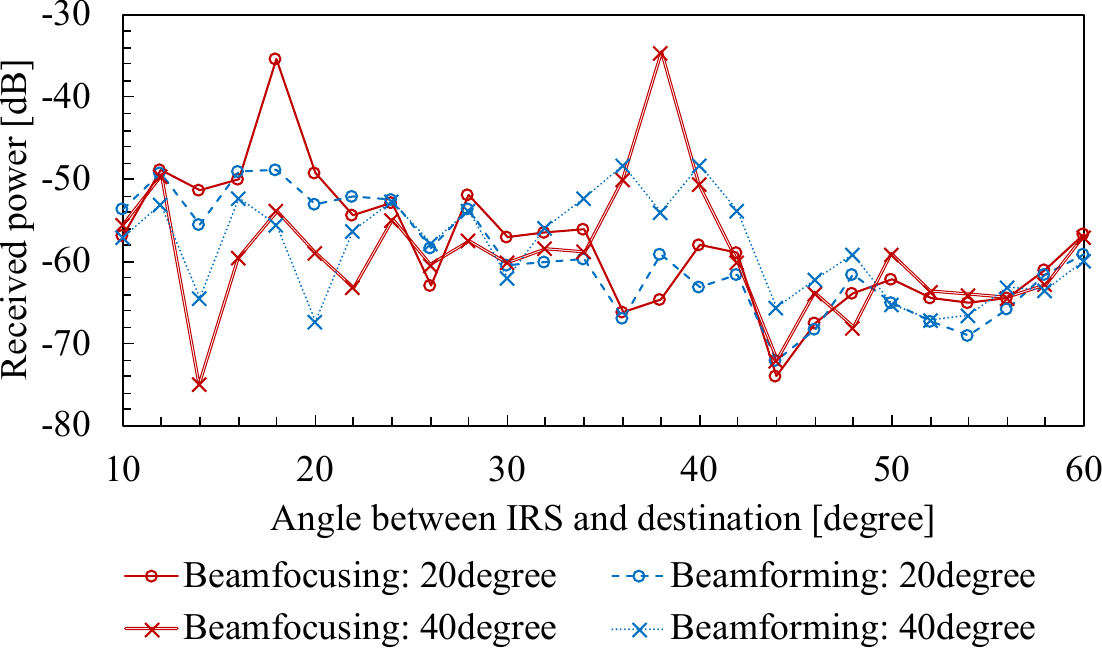}}
        \label{fig:ExResult1}
    \\
    \subfloat[Experimental results when the distance from source/destination to IRS is set to 10 m]{
        \includegraphics[width=0.9\hsize]{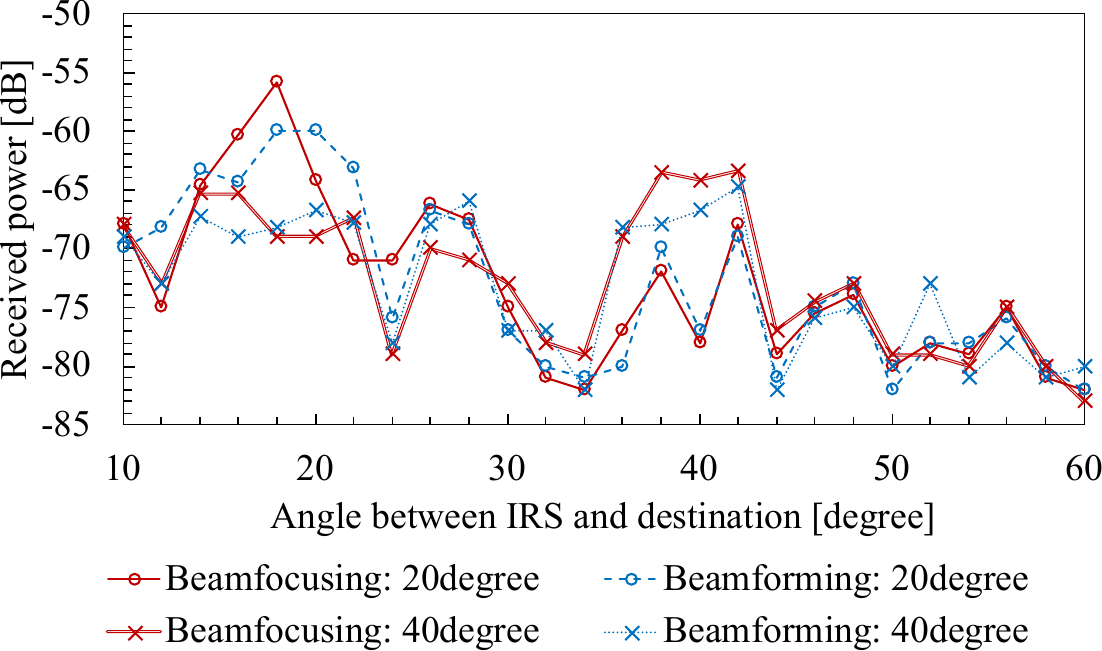}}
        \label{fig:ExResult2}
    \caption{Experimental results for beamforming and beamfocusing}
    \label{fig:ExResult}
\end{figure}
In the experiment, we used an actual IRS device with 6400 elements optimized for use at 60 GHz. The IRS phase shift was binary, either $0$ or $\pi$, and the transmission frequency was set to 60 GHz. The transmitter was positioned directly in front of the IRS with an incident angle of 0 degrees. As shown in Fig.~\ref{fig:ExEnv}, IRS control was prepared in four patterns, targeting the 20-degree and 40-degree directions for both beamforming and beamfocusing, and measurements were taken at receiver positions ranging from 10 degrees to 50 degrees. In this paper, IRS control was performed to maximize the gain at a certain angle for beamforming and at a certain point for beam focusing. Additionally, in the experiments, rounding to binary phase shifts was applied.

The measurement results with the distance set to 3 and 10 m are shown in Figs.~\ref{fig:ExResult}-(a) and ~\ref{fig:ExResult}-(b), respectively. The graphs indicate that beamfocusing provides a higher gain in the desired direction compared with beamforming. Furthermore, comparing Figs.~\ref{fig:ExResult}-(a) and ~\ref{fig:ExResult}-(b), the degradation in beamforming evidently increases as the IRS is closer because the plane-wave approximation error increases at smaller distances. If we consider an even larger IRS, the performance degradation from beamforming can be assumed to be more significant at a larger near-field, implying that a strong demand exists for beamfocusing.

\subsection{Beamtraining Overhead}
\label{subsection:Beamtraining}
Although the experiments show that beamfocusing performs better in the near-field, it also requires 3D references for positions, in contrast to beamforming that only requires 2D references. Because IRSs have multiple elements, calculating the optimum configuration for each communication path is not feasible. Therefore, a control decision method known as beamtraining is extensively considered. In beamtraining, the optimal IRS control is selected by scanning a preprepared set of IRS control patterns (codebook) to achieve the highest communication quality~\cite{Beamtraining}. However, in the near-field, positions must be scanned rather than only directions. Consequently, the number of IRS control patterns (codewords) required to achieve high communication quality increases, leading to a longer scanning time, thus increasing the overhead of control decisions. The increase in overhead reduces the time available for actual data transmission, thereby degrading communication quality. Therefore, we need a method of codebook construction with less overhead by using fewer codewords, while still satisfying the near-field requirement.

\section{Efficient Codebook Construction}
\label{section:Approaching}
In this section, recognizing the importance of codebook construction in reducing the overhead of beamtraining, we introduce different types of codebooks and clarify the type of codebook suitable for IRSs supporting near-field users.

\subsection{2D Codebook}
\label{subsection:2D}
In this study, we refer to codebooks that consider only direction---commonly used in existing methods and applicable to beamforming---as 2D codebooks. While 2D codebooks may seem entirely unsuitable for the near-field, beamforming control forms complex gain patterns in the near-field while forming gain along the beam in the far-field~\cite{Beamforming:NF}. Therefore, although degradation occurs compared with beamfocusing, they can still provide a gain to the near-field. Nonetheless, while the high gain point aligned with the target direction of the beamforming in the far-field, this does not occur in the near-field owing to the incompatible beam shape. Beamtraining can reduce the impact of this issue by scanning the codebook for the IRS control configuration with the best communication quality. Therefore, depending on the required communication quality, existing 2D codebooks can also be effective. Various methods have been proposed for 2D codebooks, such as uniform 2D codebooks that equally divide the angular range supported by the IRS and codebooks that consider changes in the beam pattern formed by the IRS depending on the angle~\cite{2D-Codebook}. It is important to note that complex gain patterns and beams formed in the far-field can potentially become sources of interference.

\subsection{3D Codebook}
\label{subsection:3D}
To provide the positional information required by beamfocusing and increase the communication quality for near-field users, a 3D codebook that divides the 3D space is required~\cite{3D-Codebook}. Despite its importance, the construction of 3D codebooks for IRS control has not been explored in detail. A simple initial method involves the uniform division of the 3D space to create a uniform 3D codebook. However, because the required control accuracy in the near-field varies with position, a uniformly divided codebook results in an enormous number of codewords required to provide high communication quality for users close to the IRS. Therefore, adopting a method that constructs a nonuniform 3D codebook is essential, which has more codewords near the IRS and fewer codewords at larger distances. Existing studies include methods such as extending the 2D codebook construction approach in the distance direction to create a 2D+1D codebook~\cite{2D+1D}. However, direction and distance are interdependent parameters, and constructing them separately can lead to performance degradation. Therefore, sufficient examination of methods for constructing non-uniform three-dimensional codebooks has not yet been conducted.

\section{Nonuniform 3D Codebook Construction}
\label{section:Proposal}
In this section, we propose a construction method for nonuniform 3D codebooks and the parameters that should be considered. Specifically, we first clarify that the cause of performance degradation in beamforming in the near-field is the plane-wave approximation error that changes nonlinearly depending on distance and angle. We also propose a method to calculate a composite control accuracy index that considers the directional patterns formed by the IRS. By referring to this control accuracy index, more efficient, nonuniform, 3D codebooks can be constructed.

\subsection{Plane-wave Approximation Error}
\label{subsection:Plane-wave}
\begin{figure}[t]
	\centering
	\includegraphics[width=1.0\hsize]{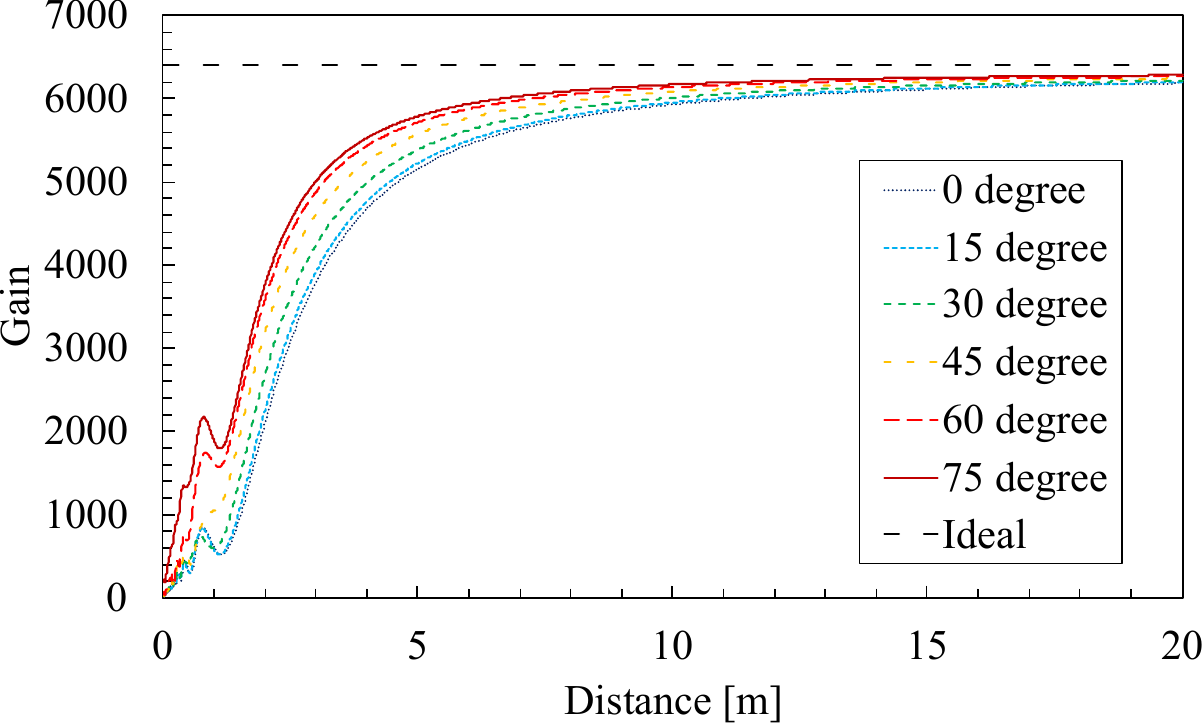}
	\caption{Plane-wave approximation error for each angle vs. distance.}
	\label{fig:PWAE}
\end{figure}
The plane-wave approximation error is the difference between the phase rotation calculated under the assumptions of plane and spherical waves. However, as IRSs have multiple elements and as phase rotation has a cyclic nature, defining a general plane-wave approximation error for a specific position on the IRS is challenging. Therefore, in this study, we visualize this by comparing the gain obtained through beamforming, which is based on the plane-wave approximation, with the maximum gain obtainable at each point. Fig.~\ref{fig:PWAE} shows the gain outcomes that can be formed through beamforming at various distances and directions (with the direction the IRS used as a reference) and compares them with the theoretical maximum gain. The assumptions include an IRS with 6400 elements and a carrier frequency of 60 GHz, similar to the IRS used in the experiment described in Section II. Thus, the theoretical maximum value of the true gain is 6400. The results indicate that the plane-wave approximation error changes nonlinearly. The wavy results for extremely short distances are due to the complex occurrence of coherent and antiphase addition when the plane-wave approximation error exceeds $\pi/2$. Additionally, the results show that the plane-wave approximation error depends on distance and angle. Based on these results, the impact on IRS control in the near-field varies with distance and angle. In areas rather distant from the IRS or along the horizontal direction on the IRS surface, the performance of beamforming approaches that of beamfocusing. As beamforming does not require scanning in the distance direction, high communication quality can be achieved with a sparse codeword arrangement in the distance direction where the beamforming performance is high. Conversely, in regions close to the IRS, where the performance of beamforming significantly decreases, arranging codewords densely is considered necessary. This indicates that simply extending the existing 2D codebook in the distance direction is insufficient, highlighting the need for a new method to construct directly nonuniform 3D codebooks.

\subsection{Control Accuracy Index}
\label{subsection:CAI}
When constructing nonuniform 3D codebooks, several factors must be considered in addition to the aforementioned plane-wave approximation error. These include the characteristics of the IRS, such as the sharper beams formed in directions closer to the front and the narrower range that can be supported along the distance and angle directions, as it gets closer to the near-field~\cite{IRS:beam}. Therefore, constructing a nonuniform 3D codebook optimized for a specific environment is a highly complex task. While one effective approach could be to organize and optimize these factors individually, this article proposes an approach that consolidates these composite factors into a control accuracy index.

\begin{figure}[t]
	\centering
	\includegraphics[width=1.0\hsize]{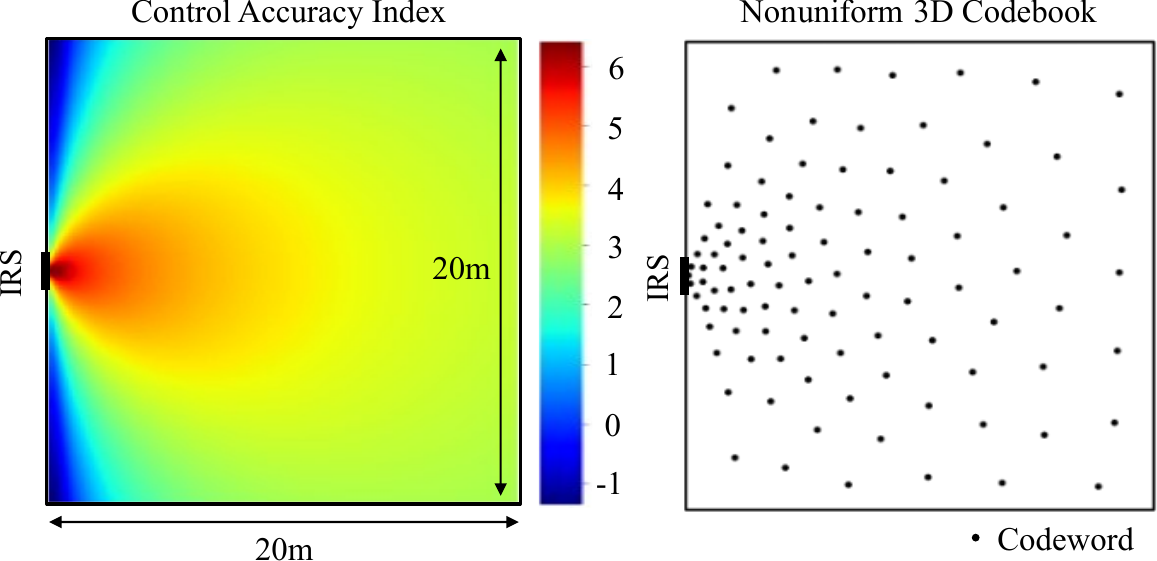}
	\caption{Control accuracy index and nonuniform three-dimensional codebook.}
	\label{fig:CAI-Codebook}
\end{figure}

The policy for constructing a nonuniform 3D codebook involves the placement of more codewords at locations requiring high-control accuracy, thus increasing the efficiency of the near-field coverage. High-control accuracy is needed in cases in which the range providing high gain with IRS control is narrow. This optimization can be redescribed as a coverage overlap minimization problem, with the constraint that the entire communication range must be covered by codewords whose coverage changes based on the control target position. To minimize the number of codewords and reduce overhead, the coverage of each codeword must be recognized, which can be inferred using the control accuracy index. This article adopts the gradient of the gain as the control accuracy index, obtained by varying the location of the signal destination after the IRS configuration is determined. For simplicity, simulations were performed to calculate the gain deterioration while the receiver was slightly moved away from the targeted location. The results for a 6400-element IRS with a 60 GHz carrier frequency are shown in Fig.~\ref{fig:CAI-Codebook}. Higher values indicate higher required control accuracy; for better understanding, the logarithms of these values are considered. The results reflect that higher control accuracy is required closer to the IRS and that fewer codewords are needed to cover areas where the beam formed by the IRS becomes broader.

Various methods can be used to generate codewords based on this index. This study uses the index as a probability distribution to generate reference points, which are then converted into codebooks using k-means clustering. Using the k-means method allows the specification of the number of codewords, enabling the construction of codebooks with a predetermined number of codewords to match the required overhead. Fig.~\ref{fig:CAI-Codebook} shows the result of constructing a codebook with 100 codewords based on the previously calculated control accuracy index. An important aspect of this method is the flexibility in designing the control accuracy index. For example, by referring to the user distribution, the placement of codewords in areas where users do not exist is possible; conversely, by setting a lower control accuracy index for positions very close to the IRS where direct waves from the base station are expected, the needs of the application can be reflected.

\section{Performance Evaluation}
\label{section:Evaluation}
In this section, we evaluate the performance of the proposed nonuniform 3D codebook and compare it to those of a uniform 3D and 2D codebooks based on simulations. The IRS was assumed to have 6400 elements, and the carrier frequency was set to 60 GHz. The evaluation procedure is as follows: first, users were randomly generated in different directions at fixed distances. Subsequently, for each user, we calculated the signal-to-noise ratio when IRS control was determined by beamtraining using each codebook. This procedure was repeated 1000 times, and the average value was output as the result for that distance. Additionally, for each codebook, we prepared versions with 100, 500, and 1000 codewords. The base station was located 15 m away from the IRS, the transmission power was set to 20 dB, and there was no direct path to the users. The channel model adopted in the simulation assumes free-space propagation, and the phase rotation and distance attenuation are calculated based on the distances between all elements of all devices. The base station and user are equipped with single antennas, and the phase control by the IRS was considered as continuous values. The evaluation metric is the signal-to-noise ratio (SNR), with noise power set at 90 dB, modeled as white Gaussian noise.

\begin{figure}[t]
	\centering
	\includegraphics[width=1.0\hsize]{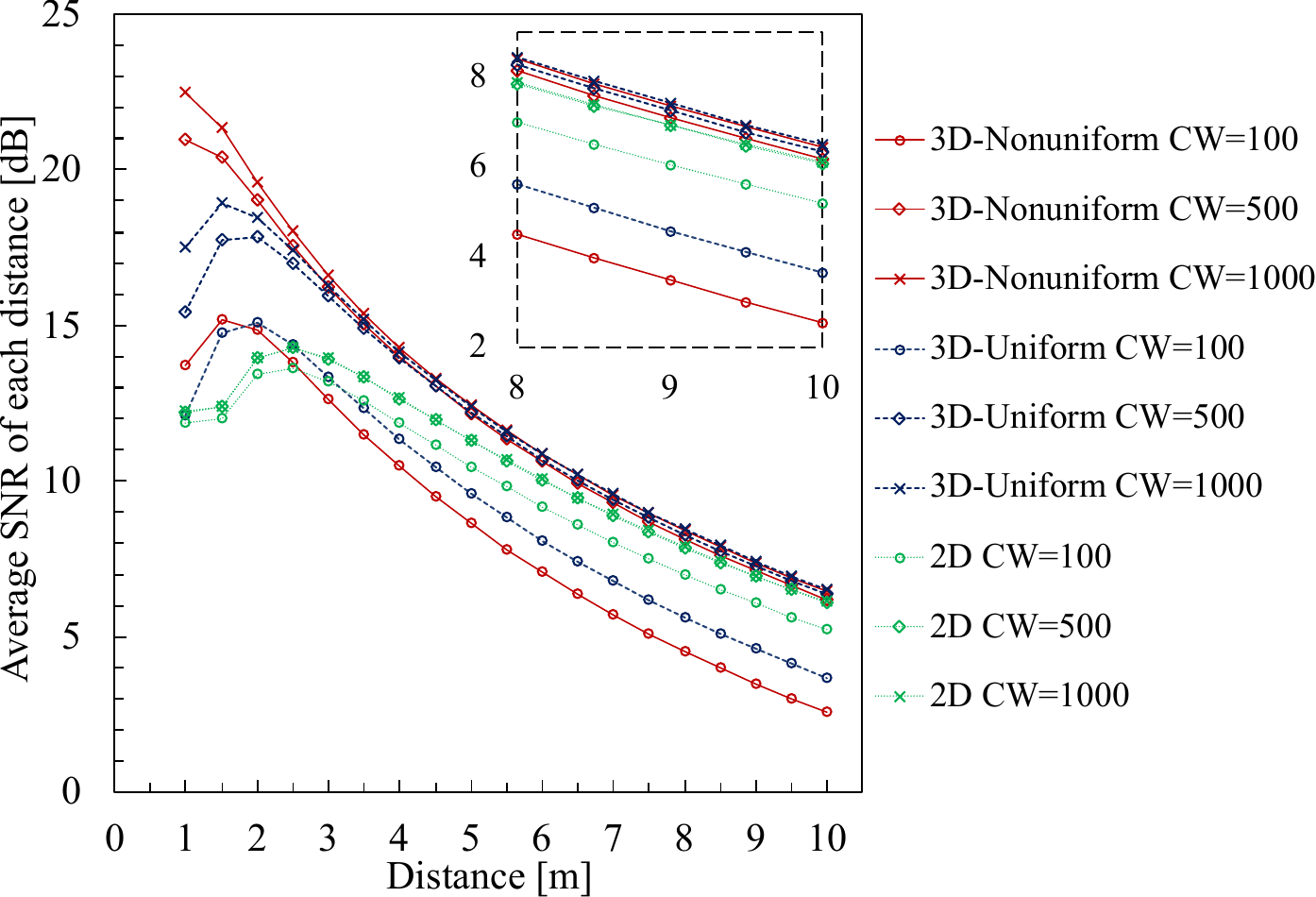}
	\caption{Average signal-to-noise ratio (SNR) plots for all codebooks as a function of distance.}
	\label{fig:SimResult}
\end{figure}

The performance evaluation results are shown in Fig~\ref{fig:SimResult}. The results indicate that the 2D codebook does not improve in performance when increasing the number of codewords from 500 to 1000, suggesting a low limit for enhancing communication performance in the near-field. However, with a small number of codewords, such as 100, the 2D codebook demonstrated high communication quality except in locations very close to the IRS. Therefore, the 2D codebook is effective when the overhead is heavily constrained, and the number of permissible codewords is limited. This is due to the IRS, with either beamforming or beamfocusing, having to service a broader area in the far-field when compared with the near-field area. As implied, beamfocusing forms a high-gain focal point, resulting in the area covered by a single IRS configuration being narrower than that of beamforming. Therefore, when the number of codewords is insufficient to cover the evaluation area, the 2D codebook, which can support a slightly wider area, becomes effective.

In the following section, we compare the uniform and nonuniform 3D codebooks. Similar to the 2D codebook results, the uniform 3D codebook---which has more IRS control configuration for the far-field---outperforms the nonuniform 3D codebook in scenarios with fewer codewords, except in locations very close to the IRS. However, as the number of codewords increases, the uniform 3D codebook distributes codewords evenly, even in areas with minimal performance improvement in the far-field, making the nonuniform 3D codebook more effective owing to its efficient codeword placement. These results indicate that the effective codebook format changes depending on the number of codewords set by the allowable overhead. Our proposed method for constructing nonuniform 3D codebooks is expected to provide further performance improvements owing to its ability to optimize the number of codewords through the flexible design of the control accuracy index. Furthermore, considering that next-generation communications are progressing with studies allowing relatively large numbers of codewords through the hierarchical execution of beamtraining, scenarios where nonuniform 3D codebooks become even more relevant will become more prominent, thereby increasing the importance of the present study~\cite{Hierarchical}.

\section{Summary and Future direction}
\label{section:Summary}
This study assessed the performance of IRS control methods in the near-field. First, practical experiments demonstrated that the extensively used beamforming approach is ineffective in the near-field, and more precise control is required through beamfocusing and its focal points. Subsequently, when implementing beamfocusing-based IRS control, scanning a 3D space is necessary compared with the existing beamtraining that scans a 2D space, resulting in a significant increase in the IRS control decision time. The study then examined the impact of codebook formats on the time required for beamtraining, introduced representative codebooks, and provided an overview of the effectiveness of 3D codebooks for the near-field. Additionally, the article proposed a construction method for nonuniform 3D codebooks, which are considered particularly effective for the near-field. Finally, the performance of the various codebooks was evaluated based on simulations, clarifying the characteristics of each codebook. These discussions highlight the importance of research on IRS control focusing on the near-field and demonstrate the effectiveness of innovative codebook construction to address several challenges that arise in the near-field. Therefore, this article clarifies the potential challenges and provides solutions for IRSs, which are highly anticipated in next-generation communications.

Future directions include leveraging the flexibility of the proposed nonuniform 3D codebook construction to develop methods adaptable to given requirements and environments, as well as extensions to multiuser and mobile communications. Additionally, while the current focus was on beamfocusing for control at a specific point, the focal point can be adjusted, thus affecting coverage. The discussion on performance evaluation indicated that control with a wider coverage is effective with a small number of codewords, suggesting the need for further examination of IRS control, which constitutes a component of the codebook. Additionally, although the control accuracy index in this paper is an indicator assuming that the coverage of codewords expands spherically, in reality, complex beam shape scaling occurs. Therefore, further performance improvements are expected through modeling that includes these factors. Finally, fast exploration methods, such as hierarchical scanning, used when applying constructed codebooks to beamtraining, are also highly effective in near-field communication, indicating the potential for methods to construct 3D codebooks hierarchically.

\section*{Acknowledgment}
This work was conducted as part of the ``Research and Development of Ultra-Large Capacity Wireless LAN using Terahertz Waves,'' supported by the Ministry of Internal Affairs and Communications (MIC), Japan.

\bibliographystyle{IEEEtran}
\bibliography{reference} %bibファイルの.bibの前の部分

% Generated by IEEEtran.bst, version: 1.14 (2015/08/26)
\begin{thebibliography}{10}
\providecommand{\url}[1]{#1}
\csname url@samestyle\endcsname
\providecommand{\newblock}{\relax}
\providecommand{\bibinfo}[2]{#2}
\providecommand{\BIBentrySTDinterwordspacing}{\spaceskip=0pt\relax}
\providecommand{\BIBentryALTinterwordstretchfactor}{4}
\providecommand{\BIBentryALTinterwordspacing}{\spaceskip=\fontdimen2\font plus
\BIBentryALTinterwordstretchfactor\fontdimen3\font minus
  \fontdimen4\font\relax}
\providecommand{\BIBforeignlanguage}[2]{{%
\expandafter\ifx\csname l@#1\endcsname\relax
\typeout{** WARNING: IEEEtran.bst: No hyphenation pattern has been}%
\typeout{** loaded for the language `#1'. Using the pattern for}%
\typeout{** the default language instead.}%
\else
\language=\csname l@#1\endcsname
\fi
#2}}
\providecommand{\BIBdecl}{\relax}
\BIBdecl

\bibitem{Demand_Chief}
X.~Sun and N.~Ansari, ``Dynamic resource caching in the {IoT} application layer
  for smart cities,'' \emph{IEEE Internet of Things Journal}, vol.~5, no.~2,
  pp. 606--613, 2018.

\bibitem{IRS_Chief}
N.~Ansari, ``A potpourri on wireless advances,'' \emph{IEEE Wireless
  Communications}, vol.~31, no.~3, pp. 4--8, 2024.

\bibitem{kato_1}
N.~Huang, C.~Dou, Y.~Wu, L.~Qian, S.~Zhou, and R.~Lu, ``Image analysis oriented
  integrated sensing and communication via intelligent reflecting surface,''
  \emph{IEEE Transactions on Cognitive Communications and Networking}, pp.
  1--1, 2024.

\bibitem{kato_2}
N.~Huang, T.~Wang, Y.~Wu, S.~Bi, L.~Qian, and B.~Lin, ``Delay minimization for
  intelligent reflecting surface assisted federated learning,'' \emph{China
  Communications}, vol.~19, no.~4, pp. 216--229, 2022.

\bibitem{IRS:Large}
E.~Bj\"{o}rnson, {\"{O}}.~\"{O}zdogan, and E.~G. Larsson, ``Intelligent
  reflecting surface versus decode-and-forward: How large surfaces are needed
  to beat relaying?'' \emph{IEEE Wireless Communications Letters}, vol.~9,
  no.~2, pp. 244--248, 2020.

\bibitem{NF:Introduction}
H.~Wymeersch, J.~He, B.~Denis, A.~Clemente, and M.~Juntti, ``Radio localization
  and mapping with reconfigurable intelligent surfaces: Challenges,
  opportunities, and research directions,'' \emph{IEEE Vehicular Technology
  Magazine}, vol.~15, no.~4, pp. 52--61, 2020.

\bibitem{Beamforming}
D.~Martinez-De-Rioja, J.~A. Encinar, E.~Martinez-De-Rioja, {\'{A}}.~F. Vaquero,
  and M.~Arrebola, ``A simple beamforming technique for intelligent reflecting
  surfaces in 5g scenarios,'' in \emph{2022 International Workshop on Antenna
  Technology (iWAT)}, 2022, pp. 249--252.

\bibitem{Beamfocusing}
S.~Droulias, G.~Stratidakis, and A.~Alexiou, ``Near-field engineering in
  {RIS}-aided links: Beamfocusing analytical performance assessment,''
  \emph{IEEE Access}, vol.~12, pp. 29\,536--29\,546, 2024.

\bibitem{Beamtraining}
R.~W. Heath, N.~González-Prelcic, S.~Rangan, W.~Roh, and A.~M. Sayeed, ``An
  overview of signal processing techniques for millimeter wave {MIMO}
  systems,'' \emph{IEEE Journal of Selected Topics in Signal Processing},
  vol.~10, no.~3, pp. 436--453, 2016.

\bibitem{Beamforming:NF}
A.~Singh, V.~Petrov, H.~Guerboukha, I.~V. Reddy, E.~W. Knightly, D.~M.
  Mittleman, and J.~M. Jornet, ``Wavefront engineering: Realizing efficient
  terahertz band communications in 6g and beyond,'' \emph{IEEE Wireless
  Communications}, vol.~31, no.~3, pp. 133--139, 2024.

\bibitem{2D-Codebook}
B.~Ning, Z.~Chen, W.~Chen, Y.~Du, and J.~Fang, ``Terahertz multi-user massive
  mimo with intelligent reflecting surface: Beam training and hybrid
  beamforming,'' \emph{IEEE Transactions on Vehicular Technology}, vol.~70,
  no.~2, pp. 1376--1393, 2021.

\bibitem{3D-Codebook}
X.~Mu, J.~Xu, Y.~Liu, and L.~Hanzo, ``Reconfigurable intelligent surface-aided
  near-field communications for 6g: Opportunities and challenges,'' \emph{IEEE
  Vehicular Technology Magazine}, vol.~19, no.~1, pp. 65--74, 2024.

\bibitem{2D+1D}
S.~Hu, H.~Wang, and M.~C. Ilter, ``Design of near-field beamforming for large
  intelligent surfaces,'' \emph{IEEE Transactions on Wireless Communications},
  vol.~23, no.~1, pp. 762--774, 2024.

\bibitem{IRS:beam}
P.~Callaghan and P.~R. Young, ``Beam- and band-width broadening of intelligent
  reflecting surfaces using elliptical phase distribution,'' \emph{IEEE
  Transactions on Antennas and Propagation}, vol.~70, no.~10, pp. 8825--8832,
  2022.

\bibitem{Hierarchical}
Y.~Yang, E.~Song, and G.~Ma, ``Hierarchical exploration based active learning
  with support vector machine,'' in \emph{2010 Second International Conference
  on Computational Intelligence and Natural Computing}, vol.~1, 2010, pp.
  299--302.

\end{thebibliography}

\begin{IEEEbiography}
[{\includegraphics[width=1 in,height=1.25 in,clip,keepaspectratio]{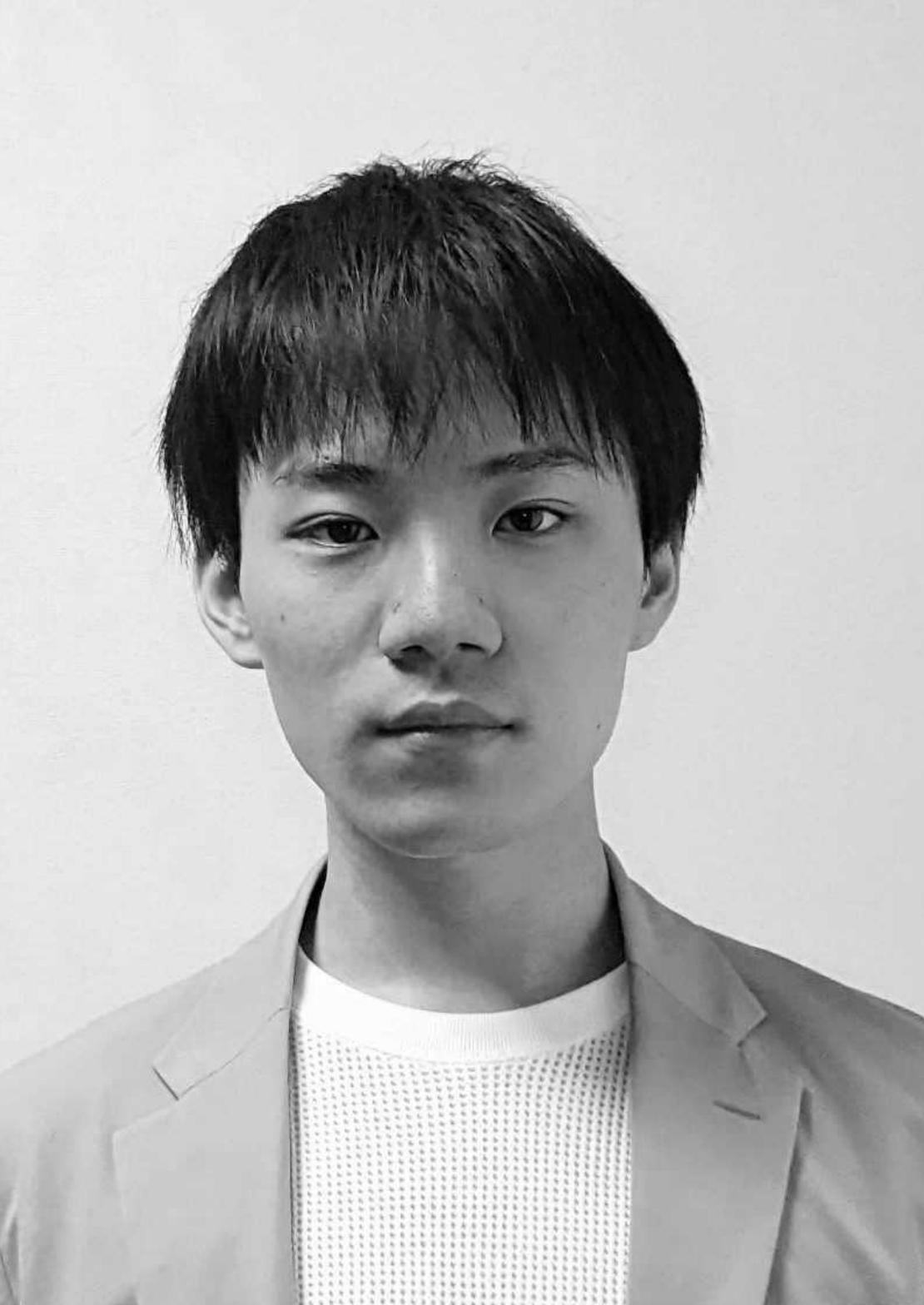}}]{Ryuhei Hibi}
(S'22) obtained his B.E. degree in Information Engineering and his M.S. degree in Information Science from Tohoku University, Sendai, Japan, in 2022 and 2024, respectively. Currently, he is pursuing his Ph.D. degree at the Graduate School of Information Sciences (GSIS) at Tohoku University. He has been a Research Fellow of the Japanese Society for the Promotion of Science since 2024. His research interests include wireless communication networks and intelligent reflective surfaces. He is a student member of the IEEE and IEICE.
\end{IEEEbiography}

\begin{IEEEbiography}[{\includegraphics[width=1in,height=1.25in,clip,keepaspectratio]{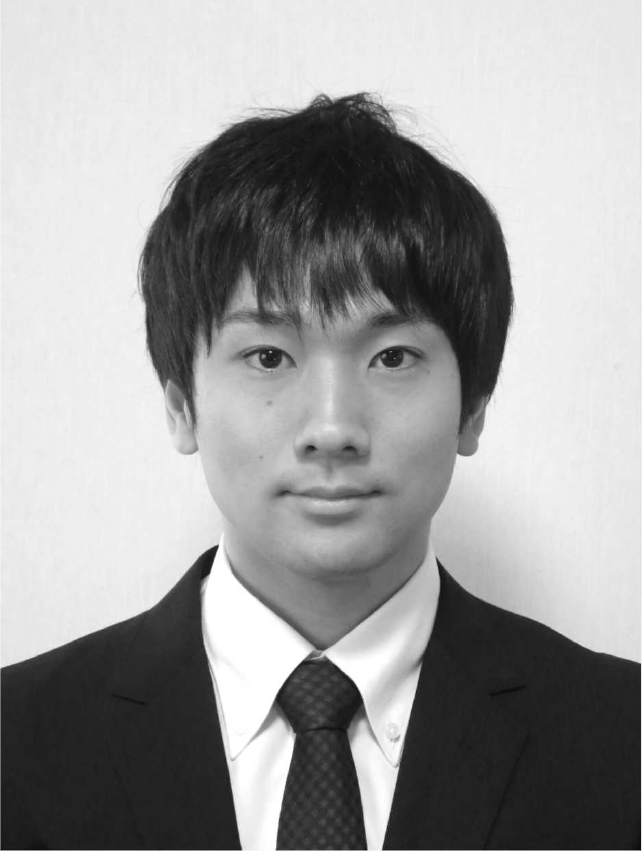}}]
{Hiroaki Hashida}
(M’24) received his Ph.D. degree from the Graduate School of Information Sciences at Tohoku University, Sendai, Japan. He is currently an Assistant Professor at the Frontier Research Institute for Interdisciplinary Sciences, Tohoku University. He was a recipient of the Presidential Award for Outstanding Students from Tohoku University and the Ikushi Prize from the Japanese Society for the Promotion of Science in 2024. His research interests include wireless communication networks and intelligent-reflecting-surface-aided wireless communication systems. He is a member of the IEEE and IEICE.
\end{IEEEbiography}

\begin{IEEEbiography}[{\includegraphics[width=1in,height=1.25in,clip,keepaspectratio]{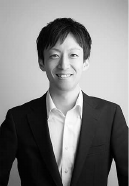}}]{Yuichi Kawamoto}
(M'16) is an Associate Professor at the Graduate School of Information Sciences (GSIS) at Tohoku University, Japan. He received his M.S. degree in 2013 and completed his Ph.D. degree in Information Science in 2016 from Tohoku University, Japan. He has published more than 60 peer-reviewed papers, including several high-quality publications in prestigious IEEE journals and conferences. His research interests cover a broad range of areas, including satellite communications, unmanned aircraft system networks, wireless and mobile networks, ad hoc and sensor networks, green networking, and network security. Moreover, he is a member of the IEEE and the Institute of Electronics, Information, and Communication Engineers (IEICE).
\end{IEEEbiography}

\begin{IEEEbiography}[{\includegraphics[width=1in,height=1.25in,clip,keepaspectratio]{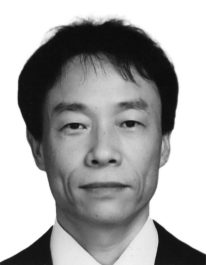}}]{Nei Kato}
(F'13) Nei Kato is a Professor and Dean of the Graduate School of Information Sciences at Tohoku University. He has studied computer networking, wireless mobile communications, satellite communications, ad hoc \& sensor \& mesh networks, UAV networks, AI, IoT, and Big Data. He has published more than 500 papers in prestigious peer-reviewed journals and conferences. He is the Editor-in-Chief of the IEEE Internet of Things Journal and Director of the Magazine of the IEEE Communications Society. He is a fellow at the Engineering Academy of Japan, IEEE, and IEICE.
\end{IEEEbiography}

\end{document}